\documentclass[sn-mathphys]{sn-jnl}

\jyear{2021}%

\theoremstyle{thmstyleone}%
\newtheorem{theorem}{Theorem}
\newtheorem{proposition}[theorem]{Proposition}%

\theoremstyle{thmstyletwo}%

\theoremstyle{thmstylethree}%

\newcommand{\ba}{\begin{equation}}
\newcommand{\ea}{\end{equation}}

\newcommand{\ZWEI}[4]{\left \{ \begin{array}{ll}
           #1 & #2 \\ & \\ #3 & #4 \end{array} \right.}
\newcommand{\DREI}[6]{\left \{ \begin{array}{ll}
           #1 & #2 \\ & \\ #3 & #4 \\ & \\ #5 & #6 \end{array} \right.}

\newcommand{\NAT} {\mathbb{N}}
\newcommand{\NATO} {\mathbb{N}_{0}}
\newcommand{\REL} {\mathbb{R}}

\begin{document}

\title[Conformity assessment of processes and lots in the framework
of JCGM 106:2012]{Conformity assessment of 
processes and lots in the framework
of JCGM 106:2012}


\author*[1]{\fnm{Rainer} 
\sur{G\"ob}}\email{goeb@mathematik.uni-wuerzburg.de}

\author[2]{\fnm{Steffen} \sur{Uhlig}}
\email{uhlig@quodata.de}
\equalcont{These authors contributed equally to this work.}

\author[3]{\fnm{Bernard} \sur{Colson}}
\email{colson@quodata.de}
\equalcont{These authors contributed equally to this work.}

\affil*[1]{\orgdiv{Department of Statistics}, \orgname{University of W\"urzburg}, \orgaddress{\street{Sanderring 2}, \city{W\"urzburg}, \postcode{D-97070}, 
\country{Germany}}}

\affil[2]{\orgdiv{QuoData GmbH}, 
\orgaddress{\street{Prellerstra\ss e 14}, \city{Dresden}, \postcode{D-01309}, 
\country{Germany}}}

\affil[3]{\orgdiv{QuoData GmbH}, 
\orgaddress{\street{Prellerstra\ss e 14}, \city{Dresden}, \postcode{D-01309}, 
\country{Germany}}}


\abstract
{ISO/IEC 17000:2020
defines conformity assessment as an "activity to determine whether specified requirements relating to a product, process, system, person or body are fulfilled". JCGM (2012) establishes a framework for accounting for measurement uncertainty in conformity assessment. The focus of JCGM (2012)  is on the conformity assessment of individual units of product based on measurements on a cardinal continuous scale. However, the scheme can also be applied to composite assessment targets like finite lots of product or manufacturing processes, and to the evaluation of characteristics in discrete cardinal or nominal scales. 

We consider the application of the JCGM scheme in the conformity assessment of finite lots or processes of 
discrete units subject to a dichotomous  quality classification as conforming and nonconforming. A lot or
process is classified as conforming if the actual proportion nonconforming does not exceed a prescribed upper tolerance limit, otherwise the lot or process is classified as nonconforming. The measurement on the lot or process is a statistical estimation of the proportion nonconforming based on attributes or variables sampling, and meassurement uncertainty is sampling uncertainty. Following JCGM (2012), we analyse the effect of measurement uncertainty (sampling uncertainty) in attributes 
sampling, and we calculate key conformity assessment parameters, in particular the producer's and consumer's risk. We suggest to integrate such parameters as a useful add-on into ISO acceptance sampling standards such as the ISO 2859 series.
}




\maketitle

\section {Introduction}
\label {Introduction}
The sampling inspection of discrete units 
of finished product assembled in
lots  is often addressed as {\em acceptance sampling}.
The ISO acceptance sampling standards, in particular the
2859 series for attributes sampling,
and 3951 series for variables sampling, provide a comprehensive and structured normative framework which is accepted
and used worldwide transversally in different industrial 
sectors.
Our study focusses on attributes sampling. Suitably adapted,
the same approach  can be used for variables sampling.

\vspace {1mm}
Sampling inspection of finished product
on a regular mass basis has been disappearing from industry
 over the past decades.
Nevertheless
acceptance sampling inspection continues
to be used
as a  tool of quality control in many 
industrial contexts (\cite{Gravesetal(2000)},
\cite{Derosetal(2008)}),
in particular in specific industrial sectors
as the pharmaceutical,
medical devices, and
food industries, 
where the use of sampling policies
is imposed by regulatory authorities.
\cite{Schilling(2017)} lists ten reasons 
for the use of acceptance sampling in modern
quality assurance.

\vspace {1mm}
The ISO 2859 attributes sampling series
evolved from the US Military 
Standard  MIL-STD-105.
MIL-STD-105A (\cite{MILSTD105A}) 
arose in 1950 as a compromise
between   sampling schemes developed 
for the US Army and US Navy 
during World War II 
(\cite{Dodge(1969a)}, \cite{Dodge(1969b)},
\cite{Dodge(1969c)}, \cite{Dodge(1969d)}, 
\cite{Schilling(2017)}).
Updated versions were issued as 
MIL-STD-105B (\cite{MILSTD105B}, 1958), 
MIL-STD-105C (\cite{MILSTD105C}, 1961), 
MIL-STD-105D
(\cite{MILSTD105D}, 1963),
and finally  MIL-STD-105E (\cite{MILSTD105E}, 1989).
The first ISO attributes sampling standard
was issued in 1974 as ISO 2859, with minor
changes adopted from  
 MIL-STD-105D. In the 1980s,
ISO 2859 was integrated into the
ISO 2859 attributes sampling series
which currently has two major 
complementary parts: 
ISO 2859-1
\cite{ISO2859-1}, widely identical 
with MIL-STD-105E (\cite{Balaetal(2008)}),
 targets a continuing series of lots
from an identifiable process,
whereas 
ISO 2859-2 \cite{ISO2859-2}
is designed for the inspection of isolated
lots (\cite{GoebandBaillie(2012)}).

\vspace {1mm}
The revisions leading from 
MIL-STD-105A to MIL-STD-105E and finally
to ISO 2859-1
were mainly on wording, arrangement 
of content, and some enlargement of the
range of quality parameters.
The core concept and the sampling tables
remained invariant.
The 
ISO 2859 attributes sampling series
widely reflects a view
of acceptance sampling shaped in 
World War II. 
Two characteristic principles of historic acceptance
sampling can be identified
(\cite{Dodge&Romig(1959)}, p.\ 11;
\cite {Duncan(1974)}, p.\ 9):
H1) The core purpose of 
acceptance sampling is the decision
on the acceptability of lots submitted
for inspection
("sorting good lots from bad", \cite{Dodge&Romig(1959)}),
  but  not 
the provision of information on the actual lot 
or process quality.
H2) Acceptance sampling 
is a stand-alone tool 
without organised embedding 
in a system of quality assurance and 
without an organised
flow of information 
to and from other quality assurance system 
components.

\vspace {1mm}
Both principles H1) and H2) are contradicting
the principles of modern quality
control systems where various 
control functions are assembled
in a coordinated system with 
an organised exchange of information,
betweeen all system components.
In such a system, 
acceptance sampling 
plays rather an adjunct than a central
role (\cite {Feigenbaum(2004)}).
Acceptance sampling schemes need to 
consider prior information
on product quality (\cite{Godfrey&Mundel(1984)}).
Beyond decision on acceptability,
the provision of information on the 
actual state of product quality 
and the 
operational risks of  sampling inspection
is a core task of acceptance sampling.

\vspace {1mm}
Conformity assessment is an important 
function of product quality control 
(\cite{Liepinaetal(2014)}).
\cite{Pendrill(2014)} defines:
\begin {quote}
Conformity assessment, broadly defined, is any activity undertaken to determine, directly or indirectly, whether an entity (product, process, system, person or body) meets relevant standards or fulfils specified requirements. 
\end {quote}
JCGM 106:2012 (\cite{JCGM(2012)}),
issued by the
Joint Committee on Guides in Metrology (JCGM),
is a guidance document for conformity
assessment under uncertain measurements of the 
target entity's relevant characteristics.
The underlying Bayesian scheme accounts
for prior information on the target 
characteristics, and allows
to calculate informative 
indicators,
in particular conformance probabilities
and risk measures. 

\vspace {1mm}
JCGM 106:2012 
focusses the conformity assessment of individual units (items) of product.
We adapt the scheme for the conformity assessment 
of finite lots or continuous processes
of discrete units
under the quality model of attributes sampling.
I.\ e., the items are
subject to a dichotomous  quality classification 
``conforming''  versus ``nonconforming'',
and the quality of the lot or process is expressed
by the proportion nonconforming.
A lot or
process is classified as conforming if the actual proportion nonconforming does not exceed a prescribed upper tolerance limit, otherwise the lot or process is classified as nonconforming. The measurement on the lot or process is a statistical estimation of the proportion nonconforming, and meassurement uncertainty is sampling uncertainty.

\vspace {1mm}
We suggest to supplement  ISO
acceptance sampling standards by 
procedures of conformity assessment of lots and processes
according to JCGM 106:2012.
This will enable the standards 
to correspond to essential 
requirements of contemporary quality control
by processing prior knowledge on lot or process
quality,  and by providing informative
conformity indicators.

\vspace {1mm}
The present study uses attributes sampling
as a paradigm for outlining the adaptation
of the JCGM 106:2012 
scheme to acceptance sampling.
Suitably modified, the approach can 
also be used fpr sampling by variables.
The study is organised in the following sections:

\section {Basics of the JCGM 106:2012 
conformity assessment scheme}
\label {Basics of the JCGM 106:2012 
conformity assessment scheme}
The 
conformity requirements
relative to some target entity are 
expressed by 
a {\em tolerance region}
({\em conformance region}) ${\cal C}$
imposed on 
the level $X$ of a scalar measurable characteristic
of the target entity. 
$X$ is addressed as the {\em target variable}.
The entity fulfills the requirements and is
conforming if $X\in {\cal C}$, 
otherwise the entity is nonconforming.
Conformity assesment proceeds by
comparing 
a measurement $Y$
 of the unknown measurand $X$ with
an {\em acceptance region} $\cal A$:
the target entity is assessed as
conforming 
(acceptance) if 
$Y\in {\cal A}$,
otherwise the target entity is 
assessed as nonconforming (rejection).
In standard applications,
both the conformance region 
${\cal C}$ and the accceptance
region ${\cal A}$ are 
one-sided or two-sided intervals.
See table \ref {TAB_QUANT_EVAL_10}
for a survey of the basic conformity assessment 
concepts.

\vspace {1mm}
JCGM 106:2012 is a Bayesian analysis scheme
and accounts for prior knowledge.
The pre-measurement 
knowledge on the measurand $X$ is
conveyed by 
the unconditional 
prior PDF $f_X$ of $X$.
The relation between the
measurement $Y$
and the measurand $X$
is uncertain,
$Y$ 
is potentially deviating
from the measurand $X$.
The post-measurement knowledge 
conveyed by $Y$
 on the measurand $X$ is expressed by the
posterior conditional PDF 
$f_{X \vert Y= y}$.
The joint unconditional PDF 
$f_{X , Y}$,
the unconditional PDFs
$f_{X}$ and $f_{Y}$, and the 
pre-measurement and post-measurement PDFs are
related by Bayes' theorem as
\ba \label {FOR_QUANT_EVAL_20}
\begin {array}{l}
\displaystyle 
f_{X \vert Y= y} (x) \;\;=\;\;
f_{Y\vert X= x}
(y)
{ f_X( x )\over 
f_{Y} 
(y) }
\;\;=\;\;
{f_{X, Y}
(x, y)
\over 
f_{Y} 
(y) }.
\end {array}
\ea

\begin{table}
\caption {Basic components of the 
JCGM 106:2012 
conformity assessment scheme}
\label {TAB_QUANT_EVAL_10}

\vspace {2mm}
\begin {center}
\begin{tabular} {p{1.0cm}|p{8.7cm}}
$X$
&  rating variable, i.\ e.,
  the level of a measurable characteristic 
used for rating the conformity of the target entity
\\[1mm] \hline  & \\[-1mm]
$Y$
&
  a measurement of the level
of the rating characteristic
\\[1mm] \hline  & \\[-1mm]
$\cal C$ &
tolerance or conformance region, i.\ e., the 
set of rating characteristic levels $X=x$ such that  
the target entity is rated as 
conforming
\\[1mm] \hline  & \\[-1mm]
$\cal A$ &
acceptance region 
of the conformity assessment test,
 i.\ e., the 
set of permissible measurements
 $Y=y$ under which 
the target entity is assessed as conforming
\\ &   \\[-1mm]  \hline  \hline
\end {tabular}
\end {center}
\end {table} 

\vspace{1mm}
JCGM 106:2012 
focusses on the continuous case where 
all PDFs are densities relative to
the Lebesgue-Borel measure.
In a straightforward  manner, the scheme
can be generalised to densities
relative to an arbitrary dominating measure
$\nu$, in particular a suitable 
counting measure in the discrete case.
To avoid an inflation of symbols,
we use $\nu$ as the generic notation for the
dominating measure in the sequel.

\section {Conformity indicators and risk
indicators suggested by the JCGM 106:2012 scheme}
\label {Conformity indicators and risk
indicators suggested by the JCGM 106:2012 scheme}

\begin{table}
\caption {Erroneous and correct results
of binary conformity assessment}
\label {TAB_QUANT_EVAL_20}

\vspace {2mm}
{\small
\begin {center}
\begin {tabular} {c|c|c}
 & $X \in {\cal C}$  & $X \notin {\cal C}$ 
\\[1mm] \hline  & & \\[-1mm]
$Y \in {\cal A}$ &
correct acceptance  & erroneous acceptance
\\[1mm] \hline  & & \\[-1mm]
$Y \notin {\cal A}$ & 
erroneous rejection & correct rejection 
\\ & &  \\[-1mm]  \hline  \hline\end {tabular}
\end {center}
}
\end {table} 

\vspace {1mm}
The suitable indicator for the conformity status
of the target entity
under a given measurement
$Y= y$
is the posterior conformance probability 
\ba \label {FOR_QUANT_EVAL_30}
\begin {array}{l}
\displaystyle 
p_{{\cal C} \vert 
 y}  \;\;=\;\; 
\mbox{P}(X\in {\cal C} \vert
Y= y)
\;\;=\;\;
\int_{y\in {\cal C}}  
f_{X \vert Y=
y} (y) \,
\mbox{d} \nu(y).
\end {array}
\ea
Erroneous and correct results of the binary
conformity assessment procedure are 
classified by table
\ref {TAB_QUANT_EVAL_20}.
The operational risks of the
conformity assessment procedure
are measured by the probabilities 
of erroneous assessment results,
with different focus for the consumer
and for the producer.

\vspace {1mm}
The consumer suffers from 
erroneously
assessing a nonconforming 
target entity as conforming. 
Thus the 
{\em specific consumer's risk}
is the conditional
probability 
\ba \label {FOR_QUANT_EVAL_40}
\begin {array}{l}
\displaystyle 
R^\star _{ \mbox{\scriptsize Con} \vert 
 y }  \;\;=\;\; 
\mbox{P} (X\notin {\cal C} \vert
Y= y
)
\;\;=\;\;
1- p_{{\cal C} \vert 
 y }
\quad \mbox{for }  y
\in {\cal A}.
\end {array}
\ea
The producer suffers from 
erroneously
assessing a conforming 
target entity as nonconforming. 
Thus the 
{\em specific producer's risk}
is the conditional
probability 
\ba \label {FOR_QUANT_EVAL_50}
\begin {array}{l}
\displaystyle 
R^\star _{ \mbox{\scriptsize Pro} \vert 
 y }  \;\;=\;\; 
\mbox{P} (X\in {\cal C} \vert
Y= y
)
\;\;=\;\;
p_{{\cal C} \vert 
 y }
\quad \mbox{for }  y
\notin {\cal A}.
\end {array}
\ea
The {\em global risks} are the unconditional
probabilities of unfavourable events
either from the consumer's or from the
producer's point of view.
The {\em global consumer's risk} is
\ba \label {FOR_QUANT_EVAL_60}
\begin {array}{l}
\displaystyle 
R _{ \mbox{\scriptsize Con}  }  \;\;=\;\; 
\mbox{P} (X\notin {\cal C}, 
Y  \in {\cal A})
\;\;=\;\;
\int_{x \notin {\cal C}, 
y \in {\cal A}}
f_{X, Y}
(x, y) \, \mbox{d}
\nu( x, y)
\;\;=\;\;
\\[4mm]
\displaystyle 
\int_{ 
y \in {\cal A}}
[1-p_{{\cal C} \vert 
 y }]
f_{Y}
( y) \, \mbox{d}
\nu(  y) ,
\end {array}
\ea
and the {\em global producer's risk} is
\ba \label {FOR_QUANT_EVAL_70}
\begin {array}{l}
\displaystyle 
R _{ \mbox{\scriptsize Pro}  }  \;\;=\;\; 
\mbox{P} (X\in {\cal C}, 
Y  \notin {\cal A})
\;\;=\;\;
\int_{x \in {\cal C}, 
y \notin {\cal A}}
f_{X, Y}
(x, y) \, \mbox{d}
\nu( x, y)
\;\;=\;\; \\[3mm]
\displaystyle 
\int_{ y \notin {\cal A}}
p_{{\cal C} \vert 
 y }
f_{Y}
( y) \, \mbox{d}
\nu(  y) .
\end {array}
\ea
From (\ref {FOR_QUANT_EVAL_60})
and (\ref {FOR_QUANT_EVAL_70}) we obtain
the equation
\ba \label {FOR_QUANT_EVAL_80}
\begin {array}{l}
\displaystyle 
R _{ \mbox{\scriptsize Con}  }  \;\;=\;\; 
\mbox{P} (Y  \in {\cal A})
-\mbox{P} (X\in {\cal C})+
R _{ \mbox{\scriptsize Pro}  }.
 \end {array}
\ea

\section {Statistical product inspection
as an instance of the JCGM 106:2012 scheme}
\label {Statistical product inspection
as an instance of the JCGM 106:2012 scheme}
We illustrate the use of the  
JCGM 106:2012 model as a conceptual scheme 
for statistical product inspection in an exemplary manner
by analysing single sampling plans
for a proportion nonconforming of a lot or a process.
Our approach can easily be adapted
to more refined sampling plans and to
other quality models, in particular to 
sampling by variables.

\vspace {1mm}
The components of the JCGM 106:2012 
conformity assessment scheme as displayed by
table \ref {TAB_QUANT_EVAL_10}
are identified as follows.
The target entity is a lot or process
of discrete units $i$ with associated
dichotomous quality indicators $U_i
\in \{0,1\}$  where $U_i =1$ if unit $i$
nonconforming, and $U_i =0$ if unit $i$
conforming.
If the target entity is the process, 
the target variable $X$ 
(rating variable) is the 
process proportion nonconforming $X=\pi$
where $\mbox{P} (U_i =1\vert \pi=p)=p$.
If the target entity is a lot of
units $i=1,...,N$,  the target variable $X$ 
(rating variable) is the 
proportion nonconforming
$(U_1+...+U_N)/N$
 of the lot, or, equivalently
the number $X=U_1+...+U_N$
of nonconforming units in the lot.
Both for the process and for the lot case, 
the tolerance region is an interval
${\cal C}= [0; x_{\cal C}]$ with 
the lower tolerance limit 0 and the upper
tolerance limit $x_{\cal C}$.

\vspace {1mm}
The measurement used for evaluating the 
unknown lot or process proportion
nonconforming $X$ is the number 
$Y$ 
of nonconforming units in a random sample of 
size $n$ from the lot or the process.
The acceptance region of acceptance sampling
by a single sampling plan $(n,c)$
is ${\cal A} = \{0,...,c\}$ with
the {\em acceptance number} $c$, i.\ e.,
the lot/process is assessed as conforming
if $Y\leq c$, and
as nonconforming if 
$Y\geq c+1$.

\vspace {1mm}
The probability distributions
of the number $Y$ 
of nonconforming units in a random sample of 
size $n$ from the lot or the process are 
analysed in sections 
\ref {Analysing process conformance} and
\ref {Analysing lot conformance},  below.

\section {Risk concepts of ISO 3534 versus risk
concepts of JCGM106:2021}
This section analyses the marked
contrast between the 
risk indicators coined by JCGM106:2021, see the above section
\ref {Conformity indicators and risk
indicators suggested by the JCGM 106:2012 scheme}, 
 and the risk indicators
defined by the ISO standard ISO 3534-2:2006 (\cite{ISO3534-2}).
A synopsis of the risk indicators from the two documents is
provided by table \ref {TAB_RISKS_JCGM_ISO_10}.

\vspace {1mm}
In both conceptual frameworks,
the specific risk indicators are conditional probabilities
where the positions of the conditioning and the conditioned
event are permuted between the frameworks.
ISO 3534-2:2006 considers the probability of an erroneous
result of the conformity assessment procedure
based on the measurement $Y$,
conditioned under a specific value of the rating characteristic
$X$.
JCGM106:2021 considers the probability of a result
of the
conformity rating based on $X$, conditioned under a specific value of the measurement $Y$.
The global risks by JCGM106:2021  are the total
 probabilities of an erroneous conformity assessment
from the consumer's and the producer's perspective.
ISO 3534-2:2006 doesn't define global risk.

\vspace {1mm}
The two risk concepts should be considered as complementary
rather than as conflicting.
In the statistical product inspection
framework established 
by paragraph 
\ref 
{Statistical product inspection
as an instance of the JCGM 106:2012 scheme}
we illustrate the complementarity 
of the risk concepts with respect
to two aspects: i) operational usefulness, and ii)
usefulness for the design of sampling plans.

\vspace {1mm}
{\bf Operational usefulness:}
The specific risk concepts by ISO are 
operationally useless for product inspection.
The value $x$
of the lot or process proportion nonconforming
in the conditioning event 
$X=x$ is unknown at the inspection operation.
Hence no inferences can be drawn from 
probabilities of the type 
$\mbox{P}(\cdot \vert X=x)$.
On the contrary, the specific risks by
JCGM106:2021 are useful for the inspection operation.
The value $Y=y$ of the number of nonconforming
units in the sample is determined in course
of the inspection operation, and can serve
for inference on the target rating variable $X$.

\vspace {1mm}
{\bf Design usefulness:}
The specific risk concepts by ISO are 
useful for the design of statistical product inspection.
Based on their insights into the respective business
processes, the producer and the consumer use to have 
precise ideas on critical levels $x$ of the
proportion nonconconforming. Thus it makes sense
 to
design inspection  procedures
in a way to restrict the  
probability of rejection at particularly
low levels of the
proportion nonconforming,
and the probability of acceptance 
at particularly high levels of the
proportion nonconforming
by specified bounds, usually 
5 \% or 10 \%, see
the standards ISO 2859-1
\cite{ISO2859-1} and 
ISO 2859-2 \cite{ISO2859-2}.
On the contrary, 
the specific risk concepts by
JCGM106:2021 are not useful for the
design  of  product inspection
procedure.
The  number
 $Y$  of nonconforming
units in a sample is not related to any
business process characteristics.
Thus designing sampling plans 
based on specific values
$Y=y$ makes no sense.

\begin{table}
\caption {Risk concepts in JCGM106:2021 and in 
ISO 3534-2:2006}
\label {TAB_RISKS_JCGM_ISO_10}

\vspace {2mm}
{\small
\begin {center}
\begin {tabular} {p{2.3cm}|p{1.0cm}|p{3.1 cm}|p{3.1cm}}
guidance document & risk type & consumer's risk & 
producer's risk
\\[1mm] \hline  & & & \\[-1mm]
 & 
specific &
$\mbox{P} ( X\notin {\cal C} \vert Y= y) $, \; 
 $y \in {\cal A} $  &
 $\mbox{P} (X\in {\cal C} \vert Y= y)$, \;
$y \notin {\cal A} $  
\\ \cline{2-4}
\smash {\raisebox{0.4cm}{JCGM106:2021}} &
 global &
$\mbox{P} ( X\notin {\cal C} ,  Y \in {\cal A} ) $  &
 $\mbox{P} (X\in {\cal C} ,  Y \notin {\cal A} )$ 
\\[1mm] \hline  & &  & \\[-1mm]
& 
specific &
$\mbox{P} ( Y \in {\cal A} \vert X=x)$ ,
\; $x\notin {\cal C} $ &
 $\mbox{P} (Y \notin {\cal A} \vert X=x)$,
\;$ x\in {\cal C}  $
\\ \cline{2-4}
\smash {\raisebox{0.4cm}{ISO 3534-2:2006}} &
 global & \multicolumn{1}{c}{---} & 
\multicolumn{1}{c}{---}
\\ & & & \\[-1mm]  \hline  \hline\end {tabular}
\end {center}
}
\end {table} 

\section {Prior information  modelling}
\label {Prior information  modelling}
Consider the statistical product inspection framework established
by section \ref {Statistical product inspection
as an instance of the JCGM 106:2012 scheme}.
As stipulated by ISO 2859-1
\cite{ISO2859-1}, the discrete items 
$..., i-1,i,i+1,...$
are the outcome of a unique and stable production
process. In stochastic terms, the series of 
dichotomous quality indicators 
$..., U_{i-1}, U_i, U_{i+1},...$
form an i.i.d.\ process of 
Bernoulli variables 
$U_i\sim Bi(1,p)$ in the 
conditional situation with
a given values $\pi=p$
of the process proportion
nonconforming.
In this framework,
prior information modelling means to specify 
the distribution of the random 
process proportion nonconforming $X=\pi$.
A finite lot of items $1,...,N$ is 
considered as a segment of the process.
Thus a prior for the 
random process proportion nonconforming 
$X=\pi$
implicitly determines a prior
for the lot proportion nonconforming, see
section \ref {Analysing lot conformance}, below.

\vspace {1mm}
As a prior for the process proportion nonconforming 
$X=\pi$
we use a beta distribution $BETA(a,b)$ defined by the PDF
\ba \label {FOR_BETA_10}
\begin {array}{l}
\displaystyle
f _{ a, b} (x) \;\; \; = \;\;\; 
\ZWEI
{ {x ^{a-1} (1-x) ^{b-1}\over B(a,b) }
 }
{ \mbox{ for }\; 0 < x <1,} 
{0} { \mbox{ otherwise, } }
\end {array}
\ea
with parameters $a, b>0$, where 
\ba \label {FOR_BETA_20}
\begin {array}{l}
\displaystyle
B( a, b)  \;\; \; = \;\;\; 
{ \Gamma (a+b) \over \Gamma (a)
     \Gamma (b) } 
\;\; \; = \;\;\; B( b, a)  
\end {array}
\ea
is the symmetric beta function
and $\Gamma$ denotes the well-known gamma function.

\vspace {1mm}
The beta distribution model has several appealing 
characteristics:
flexibility; sparse parametrisation by only two 
parameters; 
the  property of being the conjugate prior
for the binomial distribution;
the  potential to
express various density shapes like
bathtub, inverted bathtub, strictly decreasing,
strictly increasing, constant (equidistribution).
Thus the beta distribution has become 
the preferred 
distribution for expressing prior information
on random quantities ranging over compact intervals, in particular
proportions and probabilities,
in various areas of interest, e.\ g.,
interval estimation
(\cite{GoebLurz}), acceptance sampling
 (\cite {Hald}), 
audit sampling
(\cite{Godfrey&Andrews(1982)}, 
\cite{Berg(2006)}), 
risk analysis (\cite{Kendrick(2009)}),
forecasting (\cite{Kim&Reinschmid(2009)}), 
analysis of measurement error
(\cite{Eschmannetal(2019)}).

\vspace {1mm}
In industrial and business environments,
the assignment of values to the beta parameters 
$a$ and $b$ may not be based on subjective
reasoning in the sense of arbitrary 
and methodically not defensible conjectures.
In the case of repetitive sampling 
from a process with proportion nonconforming 
$\pi$,  the parameters 
of the prior information distribution can be estimated
from accumulated 
historical sample data.
If sufficient sound and reliable reference data are not available,
the features of the distribution have to be
elicited from expert opinions in interviews or panels.
The process of eliciting distprributions
from experts has received
considerable interest in the  literature, 
with particular emphasis on the beta distribution, see 
\cite{Corless(1972)}, 
\cite{Hogarth(1975)},  \cite{Kadaneetal(1980)}
\cite{Chaloner&Duncan(1983)},
\cite{O'Hagan(1998)}, \cite{Walls&Quigley(2001)}.
Software assisted approaches are considered by
\cite{Blocher&Robertson(1976)} or
\cite{Garthwaite&O'Hagan(2000)}.

\vspace {1mm}
A simple algorithm for determining the beta parameters
has the following steps:
i) Specify the support $[p_0;p_1]$ of $Y$. ii) Specify the
mean $\mu_Y$. iii) Specify a quantile $p_{\rho}$, i.\ e.,
a value $p_{\rho}$ with $F_Y(p_{\rho})=\rho$.
iv) Solve for the parameters $a$ and $b$.
In modern  industrial and  business
environments very small values 
$\pi=p$ are to be expected.
For this case, an appropriate model
is a  prior density $f_{a,b}$ decreasing on $[0;1]$
with  a large probability mass close to 0.
A simple instance of this shape is a beta distribution
$B(1.0,b)$ 
with first  parameter $a=1.0$.
Then the second parameter $b$ is uniquely determined by specifying
either the mean $\mu_Y$ or a quantile $p_{\rho}$.

\section {Analysing process conformance}
\label {Analysing process conformance}
Consider a process $..., i-1,i,i+1,...$
of discrete units $i$ with 
the dichotomous quality indicator $U_\ell$
where $U_\ell=1$ if unit
$\ell$ nonconforming, $U_\ell=0$ if unit $i$
conforming.
The target entity is the process
$(U_i)$, and the target characteristic is the 
random process proportion nonconforming
$X=\pi$. 
The tolerance region is an interval
${\cal C}= [0; x_{\cal C}]$ with 
the lower tolerance limit 0 and the upper
tolerance limit $x_{\cal C}$.
The target characteristic
(process proportion nonconforming) is evaluated by 
the number
$Y $ of nonconforming units in a random sample from the
process.
The acceptance region is  ${\cal A} = \{0,...,c\}$ with
the {\em acceptance number} $c$, i.\ e.,
the process is assessed as conforming
if $Y\leq c$, and
as nonconforming if 
$Y\geq c+1$.

\vspace {1mm}
In the process, the quality indicators $U_i$ 
are assumed to be independent Bernoulli variables,
i.\ e., conditional on a value $\pi=p$
of the process proportion nonconforming $\pi$,
each item quality indicator 
$U_i$ is distributed by the binomial distribution $Bi(1,p)$.
Hence, conditional on $\pi=p$, the number $Y$ 
of nonconforming units in a random sample of 
size $n$ from the process is distributed by the
binomial distribution $Bi(n, p)$.

\vspace {1mm}
Under the prior information model established
by section \ref {Prior information  modelling},
the random process proporion nonconforming $X=\pi$
is assumed to be distributed by a beta distribution
$BETA(a,b)$.
Hence by assertion b) of proposition
\ref {PROPO_BIVARIATE_BETABINOMIAL_10}
in the appendix 
\ref
{The univariate and the bivariate beta-binomial distributions}, 
the unconditional distribution of 
the number $Y$ 
of nonconforming units in the sample is the univariate
beta-binomial distribution 
$BETA$-$Bi(n, a,b)$ with PDF provided
by equations 
(\ref {FOR_BETABINOMIAL_10}) and
(\ref {FOR_BETABINOMIAL_20}).

\vspace {1mm}
By assertion c) of proposition
\ref {PROPO_BIVARIATE_BETABINOMIAL_10}
the posterior conditional distribution 
of the process proportion nonconforming 
$X=\pi$
under a given number  $Y=y$ 
is the  beta distribution
$BETA(a+y,
b+n-y)$
for $y \in \{0,...,n\}$.
Using the Lebesgue measure $\nu$ as
in the equations provided by section 
\ref {Conformity indicators and risk
indicators suggested by the JCGM 106:2012 scheme},
the conformity indicators and risk
indicators  can be
expressed with the CDF
$F_{ \mbox{\scriptsize 
$BETA (a+y,
b+n-y)$}} $ of the
beta distribution
$BETA(a+y,
b+n-y)$.
The conditional conformance probability is
obtained from equation (\ref {FOR_QUANT_EVAL_30}) as
\ba \label {FOR_PROCESSCONFORMITY_30}
\begin {array} {l}
\displaystyle
p_{{\cal C} \vert y }
\;\;=\;\;
\mbox{P} ( X \leq x_{\cal C} \vert
Y=y ) \;\;=\;\; 
F_{ \mbox{\scriptsize $BETA
(a+y, b+n-y)$}} 
(x _{ \cal C} ) .
\end {array}
\ea
By the equations (\ref {FOR_QUANT_EVAL_40}) and
(\ref {FOR_QUANT_EVAL_50}) 
the specific consumer's
risk and the specific producer's
risk are 
\ba \label {FOR_PROCESSCONFORMITY_40}
\begin {array}{l}
\displaystyle 
R^\star _{ \mbox{\scriptsize Con} \vert 
 y }  \;\;=\;\; 
\mbox{P} (X \notin {\cal C} \vert
Y=y )
\;\;=\;\;
\\[3mm]
\displaystyle
1- F_{
\mbox{\scriptsize $BETA
(a+y,
b+n-y )$}} 
(x _{ \cal C} ) \quad 
\mbox{for }\; y \leq c,
\end {array}
\ea
\ba \label {FOR_PROCESSCONFORMITY_50}
\begin {array}{l}
\displaystyle 
R^\star _{ \mbox{\scriptsize Pro} \vert 
 y }  \;\;=\;\; 
\mbox{P} (X\in {\cal C} \vert Y=y )
\;\;=\;\; 
\\[3mm]
\displaystyle
 F_{
\mbox{\scriptsize $BETA
(a+y, b+n-y)$}} 
(x _{ \cal C} ) \quad 
\mbox{for }\; y \geq c+1.
\end {array}
\ea
Using (\ref {FOR_QUANT_EVAL_60}), we obtain 
the global consumer's risk 
\ba \label {FOR_PROCESSCONFORMITY_60}
\begin {array}{l}
\displaystyle 
R _{ \mbox{\scriptsize Con}  }  \;\;=\;\; 
\\[3mm]
\displaystyle
\sum_{y\leq c }
{n\choose y }
 {B(a+y,
b+n-y)\over B(a,b)}
[1- F_{
\mbox{\scriptsize $BETA
(a+y,
b+n-y)$}} 
(x _{ \cal C} ) ] . 
\\[3mm]
\displaystyle
\end {array}
\ea

\section {Analysing lot conformance}
\label {Analysing lot conformance}
Consider discrete units $i$ with 
the dichotomous quality indicator $U_i$
where $U_i=1$ if unit
$i$ nonconforming, $U_i=0$ if unit $i$
conforming.
The item quality indicators
$U_1,...,U_N$ are considered
as a segment of an i.i.d.\ process
where $U_i\sim Bi(1,p)$ with the process
proportion nonconforming  $\pi = p$.

\vspace {1mm}
The target entity is a  lot 
$1,...,N$ 
of size $N$ composed of such units
with associated quality indicators
$U_1,...,U_N$.
The target characteristic is the 
number  $X= U_1+....+U_N$ of nonconforming
units in the lot, or, equivalently,
the lot proportion nonconforming $X/N$.
The tolerance region is an interval
${\cal C}= [0; x_{\cal C}]$ with 
the lower tolerance limit 0 and the upper
tolerance limit $x_{\cal C}$.
The target characteristic is evaluated by the 
number $Y$ 
of nonconforming units in a random sample of size
$n$ from the lot.
The acceptance region is  ${\cal A} = \{0,...,c\}$ with
the {\em acceptance number} $c$, i.\ e.,
the process is assessed as conforming
if $Y\leq c$, and
as nonconforming if 
$Y\geq c+1$.

\vspace {1mm}
Conditional on a value $\pi=p$ of the
process proportion nonconforming, we have  $X\sim Bi(N,p)$, 
$Y \sim Bi(n,p)$,
$X-Y \sim Bi(N-n,p)$,
where $Y$ and 
$X-Y$ are independent.
Under the prior information model established
by section \ref {Prior information  modelling},
the random process proporion nonconforming $\pi$
is assumed to be distributed by a beta distribution
$BETA(a,b)$.
Hence inserting $Z_1=Y$,$n_1=n$, $Z_2=X-Y$,
$n_2= N-n$ in the assertion c) of proposition
\ref {PROPO_BIVARIATE_BETABINOMIAL_10}
in the appendix 
\ref
{The univariate and the bivariate beta-binomial distributions}
we obtain
for $y=0,...,n$, $x=y, ..., N-n$
\ba \label {FOR_BETALOTANALYSIS_40}
\begin {array} {l}
\displaystyle
f_{X \vert Y= y} (x) \;\;=\;\;
\mbox{P} (X-Y=x-y\vert Y=y) \;\;=\;\;
\\[6mm]
\displaystyle
{N-n\choose x- y } {B( a+ x, b+N-x)
\over
B(a+y, b+n-y) }
\;\;=\;\; f_{N-n, a+y,b+n-y}( x-y)
\end {array}
\ea
where the PDF $f_{N-n, a+y,b+n-y}$
of the  beta-binomial distribution 
$BETA$-$Bi(N-n, a+y,b+n-y)$ is defined by 
equation (\ref {FOR_BETABINOMIAL_10}).
By assertion b) of proposition
\ref {PROPO_BIVARIATE_BETABINOMIAL_10}
in the appendix 
\ref
{The univariate and the bivariate beta-binomial distributions}
the unconditional distrinution of the number $Y$
of nonconforming units in the sample is the 
beta-binomial distribution $BETA$-$Bi(n,a,b)$.

\vspace {1mm}
Using the Lebesgue measure $\nu$ as
in the equations provided by section 
\ref {Conformity indicators and risk
indicators suggested by the JCGM 106:2012 scheme},
the conformity indicators and risk
indicators  can be
expressed with the CDF
$F_{ \mbox{\scriptsize 
$BETA$-$Bi(N-n, a+y, b+n-y)$}} $ of the
beta-binomial distribution
$BETA$-$Bi(N-n,a+y, b+n-y)$.
The conditional conformance probability is
obtained from equation (\ref {FOR_QUANT_EVAL_30}) as
\ba \label {FOR_LOTCONFORMITY_30}
\begin {array} {l}
\displaystyle
p_{{\cal C} \vert y }
\;\;=\;\;
\mbox{P} ( X \leq x_{\cal C} \vert
Y=y ) \;\;=\;\; 
F_{ \mbox{\scriptsize $BETA-Bi
(N-n, a+y, b+n-y)$}} 
(x _{ \cal C} -y) .
\end {array}
\ea
By the equations (\ref {FOR_QUANT_EVAL_40}) and
(\ref {FOR_QUANT_EVAL_50}) 
the specific consumer's
risk and the specific producer's
risk are 
\ba \label {FOR_LOTONFORMITY_40}
\begin {array}{l}
\displaystyle 
R^\star _{ \mbox{\scriptsize Con} \vert 
 y }  \;\;=\;\; 
\mbox{P} (X \notin {\cal C} \vert
Y=y )
\;\;=\;\;
\\[3mm]
\displaystyle
1- F_{
\mbox{\scriptsize $BETA$-$Bi
(N-n, a+y,
b+n-y )$}} 
(x _{ \cal C}-y ) \quad 
\mbox{for }\; y \leq c,
\end {array}
\ea
\ba \label {FOR_LOTFORMITY_50}
\begin {array}{l}
\displaystyle 
R^\star _{ \mbox{\scriptsize Pro} \vert 
 y }  \;\;=\;\; 
\mbox{P} (X\in {\cal C} \vert Y=y )
\;\;=\;\; 
\\[3mm]
\displaystyle
 F_{
\mbox{\scriptsize $BETA-Bi
(N-n,a+y, b+n-y)$}} 
(x _{ \cal C}-y ) \quad 
\mbox{for }\; y \geq c+1.
\end {array}
\ea
Using (\ref {FOR_QUANT_EVAL_60}), we obtain 
the global consumer's risk 
\ba \label {FOR_LOTCONFORMITY_60}
\begin {array}{l}
\displaystyle 
R _{ \mbox{\scriptsize Con}  }  \;\;=\;\; 
\\[3mm]
\displaystyle
\sum_{y\leq c }
{n\choose y }
 {B(a+y,
b+n-y)\over B(a,b)}
[1- F_{
\mbox{\scriptsize $BETA-Bi
(N-n, a+y,
b+n-y)$}} 
(x _{ \cal C} ) -y ] . 
\\[3mm]
\displaystyle
\end {array}
\ea

\section {Numerical results on lot conformance indicators}
The consumer's direct concern is the 
respective lot of product under consideration.
The process is relevant for long-term considerations
whereas
the state of the lot direct affectly affects the 
consumer's technical and business process.
Hence it is most interesting to
 consider the conformance indicators  
generally defined by section
\ref {Conformity indicators and risk
indicators suggested by the JCGM 106:2012 scheme}
for the case of lot evaluation,
using the analytical results of section 
\ref {Analysing lot conformance}.

\vspace {1mm}
For inspecting for a proportion of
nonconforming items,
ISO 2859-1
\cite{ISO2859-1}
is the most common sampling standard in industry.
The standard targets the process behind the stream
of lots. However, in business practice the consumer 
needs to take decisions  on
each individual lot submitted for inspection.
So the indvidual lot is the primary target
for conformity assessment.

\vspace {1mm}
We consider the sampling plans
from
ISO 2859-1 in the most widely used categories:
 normal inspection, general inspection level II.
ISO 2859-1 indexes the sampling plans in the lot size
$N$  and in 
the AQL (acceptance quality limit) which is defined 
as the ``worst tolerable quality level''
(ISO 3534-2:2006 \cite{ISO3534-2}),
here the worst tolerable level of the 
proportion nonconforming.
As such, the AQL is the natural choice for the 
upper bound  of the 
tolerance or conformance region
$\cal C$, see table \ref {TAB_QUANT_EVAL_10}.
ISO 2859-1 lists  26  AQL percentages 
in approximate geometric progression
$0.01 \times 10^{i/5}$, $i=0,..., 25$.
The 7 largest values are inadequate 
for a percentage nonconforming,
and do only make sense for
quality statements in
the case of nonconformities per 100 items.
For our numerical experiments we consider the
lot size $N=1200$.
So we restrict attention to the 19 AQL values 
and the corresponding single sampling plans from ISO 2859-1
under lot size $N=1200$
for  normal inspection, general inspection level II,
listed by table \ref
{TAB_AQL_SAMPLINGPLANS_10}.

\vspace {1mm}
So as to adapt the AQL  to 
the lot quality scheme exposed by
section \ref {Statistical product inspection
as an instance of the JCGM 106:2012 scheme}
we have
to round the AQL to a value which can appear
as a fraction $X/N$ for some number
$X$ of nonconforming units
under a given lot size $N$.
In agreement with the definition as the
``worst tolerable quality level''
we use the value
$x_{\cal C}= \lfloor 
\mbox{AQL}/100 \times N\rfloor$ as the 
upper bound for the tolerance or conformance 
region for the number $X$ of 
nonconforming units in the lot.

\vspace {1mm}
Prior information is expressed
by a beta distribution 
of the process proportion nonconforming, see section
\ref {Prior information  modelling}. We evaluate
the effect of prior information by considering
the beta distributions $BETA(a,b)$ 
with the parameter pairs $(a,b)$
listed by table \ref {TAB_BETAPAR_10}.
As visible from the respective means and 99 \% quantiles 
the parameter pairs cover cover a broad range of 
states of the process proportion nonconforming.
In the light of the quality levels usually required
in modern industrial and business environments
all considered distributions amount to relatively conservative
assumptions.
Even the most restrictive assumption with
$a= 0.24$ and $b= 78.12$, mean 0.003 
(3 permil nonconorming units on
average) and 99 \% quantile $z_{a,b}(0.99)= 0.030 $
(percentage of nonconforming units not exceeding 3 percent
in 99 \% of all cases) is not unrealistically optimistic
in a modern industrial environment.
The beta distribution with $(a,b)=(1,1)$ amounts to equidistribution
of the proportion nonconforming over the
unit interval, i.\ e., no specific prior information.

\begin{figure}
\begin{center}
\caption {Lot conformance probability
$p_{{\cal C}\vert 0}$ 
for lot conformance limit $x_{\cal C}= \lfloor 
\mbox{AQL}/100 \times N\rfloor$,
 under $x_{ \mbox{\scriptsize ev}}=0$
nonconforming units in sample from lot of size 
$N=1200$, $a=1.00=b$ (blue),
$a=0.57$, $b=37.67$ (orange).
Sample sizes from
ISO 2859-1, normal inspection, general inspection level II, see
table
\ref {TAB_AQL_SAMPLINGPLANS_10}.} 
\label  {GRA_PROCESSCONFORMPROB_10} 
{\small
\scalebox {1.0}
 {\includegraphics* [0, 0] [300, 200]
{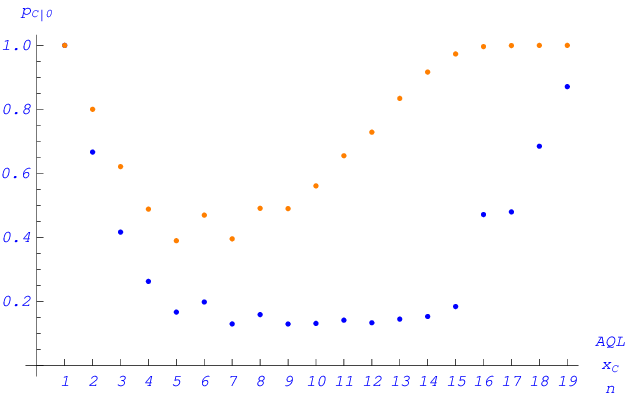}}
}
\end{center}
\end{figure}

\begin{figure}
\begin{center}
\caption {Lot conformance probability
$p_{{\cal C}\vert 0}$ 
for lot conformance limit $x_{\cal C}= \lfloor 
\mbox{AQL}/100 \times N\rfloor$,
$\mbox{AQL}=0.04$ in percent,
under $x_{ \mbox{\scriptsize ev}}=0$
nonconforming units in sample
of size $n=315$ from lot of size 
$N=1200$ for the nine priors
$BETA(a,b)$ from table
\ref {TAB_BETAPAR_10}.
Sample size from
ISO 2859-1, normal inspection, general inspection level II,
$\mbox{AQL}=0.04$ in percent.} 
\label  {GRA_CONFORMPROB_30}
 {\small
\scalebox {1.0}
 {\includegraphics* [0, 0] [300, 200]
{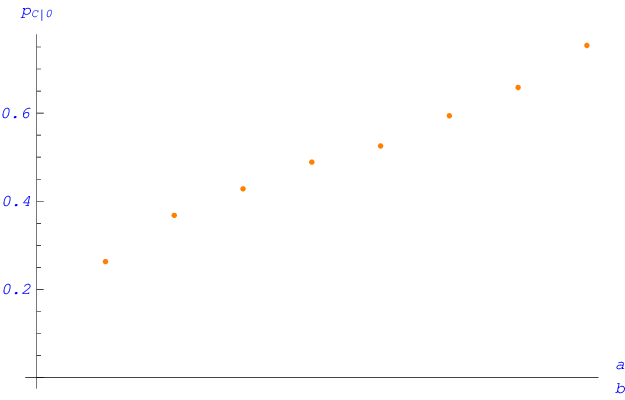}}
}
\end{center}
\end{figure}

\vspace {1mm}
The figures \ref {GRA_PROCESSCONFORMPROB_10}  and 
\ref {GRA_CONFORMPROB_30} provide the
lot conformance probability
$p_{{\cal C}\vert 0}$ under $x_{ \mbox{\scriptsize ev}}=0$
observed nonconforming units in a sample
 from a lot of size 
$N=1200$.
Figure \ref {GRA_PROCESSCONFORMPROB_10} compares the
effect of the two extreme 
priors $(a,b)=(1,1)$ (no prior information)
and $(a,b)= (0.24 , 78.12)$, 
under the 19 AQLs from table
\ref {TAB_AQL_SAMPLINGPLANS_10}.
Without specific 
prior information, the 
conformance probability is particularly small
under the sample sizes corresponding
to moderate AQLs. For the latter, 
specific prior information is able 
to leverage the 
conformance probability to moderate up to high
values.
Figure \ref {GRA_CONFORMPROB_30} illustrates
the leveraging effect of prior 
information along the nine 
pairs $(a,b)$ from table 
\ref {TAB_BETAPAR_10}
under the specific choice
$\mbox{AQL}=0.04$ in percent which amounts to
a sample size $n=315$.

\begin{figure}
\begin{center}
\caption {Specific consumer's risk 
$R^\star _{ \mbox{\scriptsize Con} \vert 
Y =c } $ and 
specific producer's risk
$R^\star _{ \mbox{\scriptsize Pro} \vert 
Y =c+1 } $,
under the 
lot conformance limit $x_{\cal C}= \lfloor 
\mbox{AQL}/100 \times N\rfloor$,
lot size 
$N=1200$, $a=1.00$, $b=1.00$.
AQLs and
sampling plans from
ISO 2859-1, normal inspection, general inspection level II, see
table
\ref {TAB_AQL_SAMPLINGPLANS_10}.} 
\label  {GRA_SPECIFICRISK_10}
 {\small
\scalebox {0.75}
 {\includegraphics* [0, 0] [300, 200]
{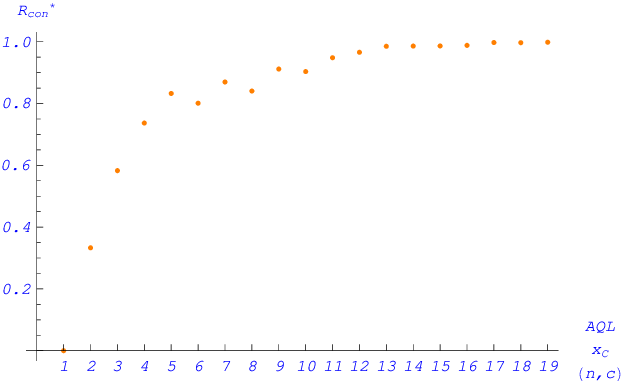}}
\\[3mm]
\scalebox {0.75}
 {\includegraphics* [0, 0] [300, 200]
{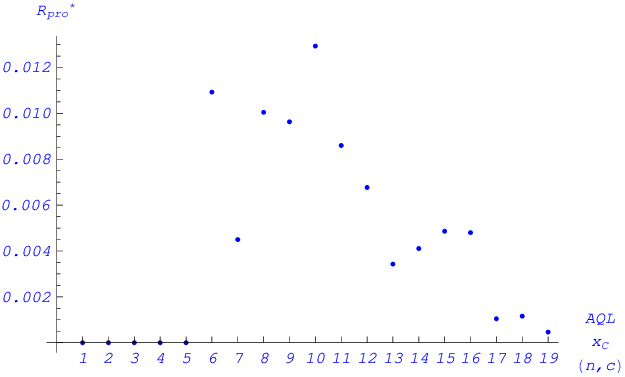}}
}
\end{center}
\end{figure}

\begin{figure}
\begin{center}
\caption {Specific consumer's risk 
$R^\star _{ \mbox{\scriptsize Con} \vert 
Y =c } $ and 
specific producer's risk
$R^\star _{ \mbox{\scriptsize Pro} \vert 
Y =c+1 } $,
lot conformance limit $x_{\cal C}= \lfloor 
\mbox{AQL}/100 \times N\rfloor$,
lot size 
$N=1200$, $a=0.57$, $b=37.67$.
AQLs and
sampling plans from
ISO 2859-1, normal inspection, general inspection level II, see
table
\ref {TAB_AQL_SAMPLINGPLANS_10}.} 
\label  {GRA_SPECIFICRISK_20}
 {\small
\scalebox {0.75}
 {\includegraphics* [0, 0] [300, 200]
{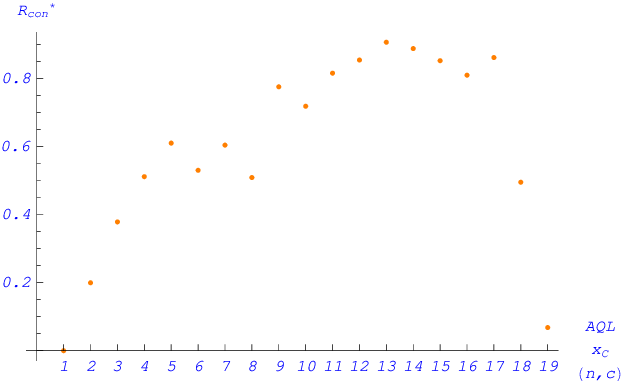}}
\\[3mm]
\scalebox {0.75}
 {\includegraphics* [0, 0] [300, 200]
{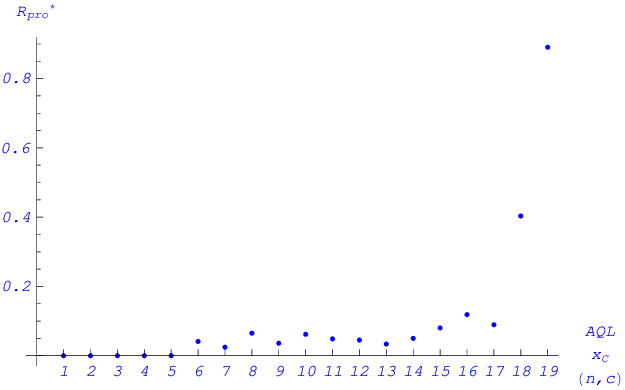}}
}
\end{center}
\end{figure}

\vspace {1mm}
The figures 
\ref {GRA_SPECIFICRISK_10} and
\ref {GRA_SPECIFICRISK_20}
consider the specific risk indicators
introduced in section \ref
{Conformity indicators and risk
indicators suggested by the JCGM 106:2012 scheme}:
i) the  specific consumer's risk defined
by formula (\ref {FOR_QUANT_EVAL_40})
as the 
conditional probability of finding an
actually nonconforming lot 
with $X> x_{\cal C}$  conditional 
on a sample result which enforces
assessment of the lot as conforming;
ii) the  specific producer's risk defined
by formula (\ref {FOR_QUANT_EVAL_50})
as the 
conditional probability of finding an
actually conforming lot conditional 
on a sample result which enforces
assessment of the lot as nonconforming.

\vspace {1mm}
By the model of section
\ref {Statistical product inspection
as an instance of the JCGM 106:2012 scheme}, 
under a single sampling plan $(n,c)$ sample 
results $0\leq Y\leq c$ 
(number $Y$ of nonconforming units
in sample not exceeding the acceptance
number $c$) lead to assessing
the lot as conforming, and sample results
$Y\geq c+1$ 
(number $Y$ of nonconforming units
in sample exceeding the acceptance
number $c$) lead to assessing the lot as nonconforming.
In the  figures 
\ref {GRA_SPECIFICRISK_10} and
\ref {GRA_SPECIFICRISK_20}
we consider sample results at the margin of the acceptance
region: $Y=c$ as the condition
for the consumer's risk
$ R^\star _{ \mbox{\scriptsize Con} \vert 
 Y =c } = \mbox{P} (X > x_{\cal C} \vert
Y= c) $, and
$Y=c+1$ as the condition
for the producer's risk
$ R^\star _{ \mbox{\scriptsize Pro} \vert 
 Y =c+1 } = \mbox{P} (X \leq x_{\cal C} \vert
Y= c+1) $.
Figure \ref {GRA_SPECIFICRISK_10} considers the
beta prior with  $(a,b)=(1,1)$ (no prior information),
whereas figure \ref {GRA_SPECIFICRISK_20}
considers the most restrictive prior with 
$(a,b)= (0.24 , 78.12)$.  
The calculations range over the 
19 AQLs with corresponding conformance limit
$x_{\cal C}$ and single 
sampling plans $(n,c)$ listed  by table
\ref {TAB_AQL_SAMPLINGPLANS_10}.

\vspace {1mm}
The figures \ref {GRA_SPECIFICRISK_10} and
\ref {GRA_SPECIFICRISK_20}
show that 
under both priors, the producer's risk is 
low and the consumer's risk is high.
Particularly under the small sample sizes 
recommended by ISO 2859-1 for large AQL, the
consumer's risk is close to 1.
In the considered situation of $Y=c$
nonconforming units in the sample, so that the
lot is marginally accepted, it is 
practically certain that the consumer will
erroneously 
assess a nonconforming lot with a proportion
nonconforming exceeding the AQL as conforming.
In contrast, 
in the considered situation of $Y=c+1$
nonconforming units in the sample, so that the
lot is marginally rejected, it is 
fairly uncertain that the producer will
erroneously 
assess a conforming lot with a proportion
nonconforming below the AQL as nonconforming.
It is intuitively plausible that the 
restrictive prior with 
$(a,b)= (0.24 , 78.12)$ reduces the 
consumer's risk and slightly increases the 
producer's risk since good quality is more
likely under $(a,b)= (0.24 , 78.12)$ than under
$(a,b)= (1,1$.

\begin{figure}
\begin{center}
\caption {Global consumer's risk 
$R_{ \mbox{\scriptsize Con} } $ and 
global producer's risk
$R _{ \mbox{\scriptsize Pro} } $,
lot conformance limit $x_{\cal C}= \lfloor 
\mbox{AQL}/100 \times N\rfloor$,
lot size 
$N=1200$, $a=1.00$, $b=1.00$.
AQLs and
sampling plans from
ISO 2859-1, normal inspection, general inspection level II, see
table
\ref {TAB_AQL_SAMPLINGPLANS_10}.} 
\label  {GRA_GLOBALRISK_10}
 {\small
\scalebox {0.75}
 {\includegraphics* [0, 0] [300, 200]
{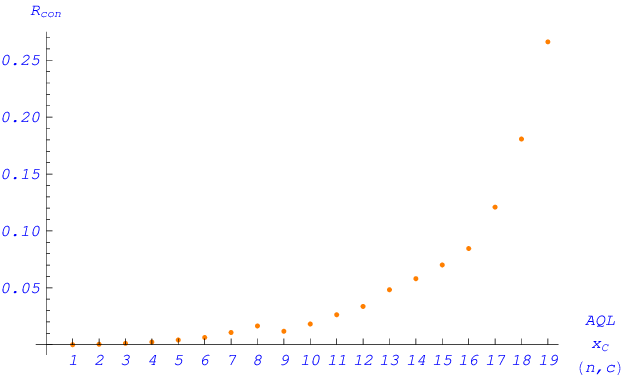}}
\\[3mm]
\scalebox {0.75}
 {\includegraphics* [0, 0] [300, 200]
{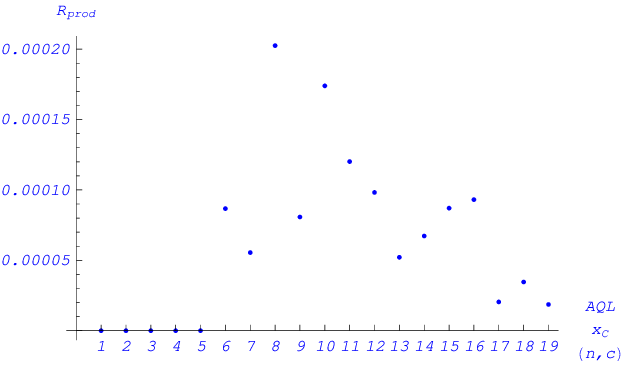}}
}
\end{center}
\end{figure}

\begin{figure}
\begin{center}
\caption {Global consumer's risk 
$R_{ \mbox{\scriptsize Con} } $ and 
global producer's risk
$R _{ \mbox{\scriptsize Pro} } $,
lot conformance limit $x_{\cal C}= \lfloor 
\mbox{AQL}/100 \times N\rfloor$,
lot size 
$N=1200$, $a=0.57$, $b=37.67$.
AQLs and
sampling plans from
ISO 2859-1, normal inspection, general inspection level II, see
table
\ref {TAB_AQL_SAMPLINGPLANS_10}.} 
\label  {GRA_GLOBALRISK_20}
 {\small
\scalebox {0.75}
 {\includegraphics* [0, 0] [300, 200]
{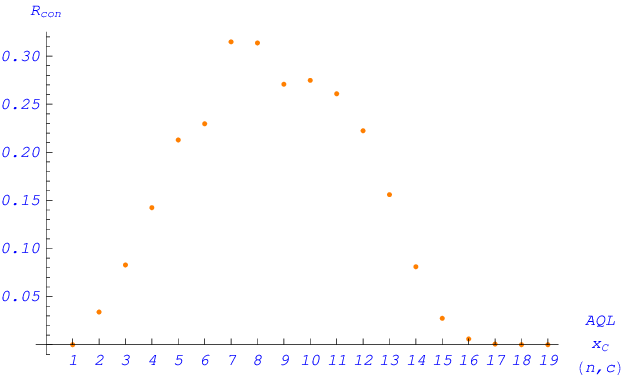}}
\\[3mm]
\scalebox {0.75}
 {\includegraphics* [0, 0] [300, 200]
{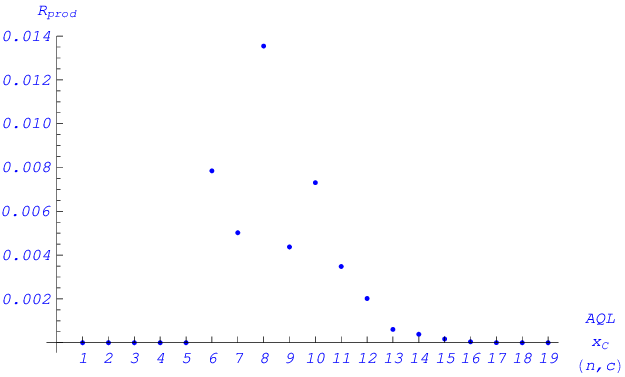}}
}
\end{center}
\end{figure}

\vspace {1mm}
The figures \ref {GRA_GLOBALRISK_10} and 
\ref {GRA_GLOBALRISK_20} illustrate the 
global risks defined in paragraph 
\ref {Conformity indicators and risk
indicators suggested by the JCGM 106:2012 scheme}
by 
(\ref {FOR_QUANT_EVAL_60}) and (\ref
{FOR_QUANT_EVAL_70}).
The dependence on specific sampling results
 $Y=y$ chosen as $Y=c$ (upper margin of acceptance
region) for figure \ref {GRA_SPECIFICRISK_10} 
and as $Y=c+1$ (lower margin of rejectio
region) for figure \ref {GRA_SPECIFICRISK_20} 
vanishes 
in the global risks.

\begin{table}
\caption {AQL values (in percent) from ISO 2859-1 
with corresponding upper conformance 
limits $x_{\cal C}= \lfloor 
\mbox{AQL}/100 \times N\rfloor$
for the number $X$ of 
nonconforming units in the lot, and 
single sampling plans $(n,c)$
from ISO 2959-1
for  normal inspection, general inspection level II.
}
\label {TAB_AQL_SAMPLINGPLANS_10}

\vspace {2mm}
{\small
\begin {center}
\begin {tabular}{r|rrrr}
&AQL &  $x_{\cal C}$ & $n$ & $c$\cr
\hline
1& 0.010 & 0 & 1200 & 0\cr  
2& 0.015 & 0 & 800 & 0\cr  
3& 0.025 & 0 & 500 & 0\cr  
4& 0.040 & 0 &   315 & 0\cr 
5&  0.065 & 0 & 200 & 0\cr  
6& 0.100 & 1 & 125 & 0\cr  
7& 0.150 & 1 & 80 &   0\cr  
8& 0.250 & 3 & 50 & 0\cr  
9& 0.400 & 4 & 125 & 1\cr  
10& 0.650 & 7 & 80 & 1\cr  
11& 1.000 & 12 &   80 & 2\cr  
12& 1.500 & 18 & 80 & 3\cr  
13& 2.500 & 30 & 80 & 5\cr  
14& 4.000 & 48 & 80 & 7\cr  
15& 6.500 &   78 & 80 & 10\cr  
16& 10.000 & 120 & 80 & 14\cr  
17& 15.000 & 180 & 80 & 21\cr  
18& 25.000 & 300 & 50 &   21\cr  
19& 40.000 & 480 & 32 & 21
\end {tabular}
\end {center}
}
\end {table} 
\begin{table}
\caption {Representative pairs $(a,b)$ of parameters of the
beta distribution $BETA(a,b)$ with 
the corresponding mean and 99 \% quantile
$z_{a,b}(0.99)$.}
\label {TAB_BETAPAR_10}

\vspace {4mm}
{\footnotesize
\begin {center}
\begin{tabular}{c|c|c|c}
$a$ & $b$ & mean & $z_{a,b}(0.99)$ 
\\[1mm]  \hline  & & & \\[-1mm] 
				1.00    & 1.00     & 0.500 & 0.990  
\\[1mm]  \hline  & & & \\[-1mm] 
				0.78 & 25.21 & 0.030 & 0.150 
\\[1mm]  \hline  & & & \\[-1mm] 
				0.67 & 32.67 & 0.020 & 0.110 
\\[1mm]  \hline  & & & \\[-1mm] 
				0.57 & 37.67 & 0.015 & 0.090 
\\[1mm]  \hline  & & & \\[-1mm] 
				0.52 & 46.79 & 0.011 & 0.070 
\\[1mm]  \hline  & & & \\[-1mm] 
				0.43 & 60.46 & 0.007 & 0.050 
\\[1mm]  \hline  & & & \\[-1mm] 
				0.35 & 69.50 & 0.005 & 0.040 
\\[1mm]  \hline  & & & \\[-1mm] 
				0.24 & 78.12 & 0.003 & 0.030 
\\   & & & \\[-1mm] \hline  \hline
			\end{tabular}
\end {center}
}
\end {table} 

\vspace {1mm}

\clearpage
\section {Conclusion and outlook}
The present paper provides a basic model for the
application of the JCGM conformity assessment scheme
in acceptance sampling.
The mathematical-statistical methodology is developed 
for attributes sampling.
In the next step, the  methodology needs to be adapted
to variables sampling.


\clearpage

\begin{appendices}

\section
{The univariate and the bivariate beta-binomial distributions}
\label 
{The univariate and the bivariate beta-binomial distributions}
The univariate discrete distribution  defined by the
PDF
\ba \label {FOR_BETABINOMIAL_10}
\begin {array} {l}
\displaystyle 
f_{n,a,b}(z) \;\;=\;\; 
{B(z+a, n-z+b)  \over B(a,b)} {n\choose z} 
\quad \mbox{for $z=0,...,n$}
\end {array}
\ea
is called {\em beta-binomial
distribution} 
with parameters $a,b \in (0;+\infty)$ and $n \in \NAT$,
in short {\em beta-binomial distribution 
$BETA$-$Bi(n, a,b)$}.
Using the equations
(\ref {FOR_BINOMKOEFF_15}) and
(\ref {FOR_BETAFUNKTION_45}) provided
by the appendix 
\ref {Facts on binomial coefficients
and the beta function}, the PDF 
(\ref {FOR_BETABINOMIAL_10}) can be expressed  
as
\ba \label {FOR_BETABINOMIAL_20}
\begin {array} {l}
\displaystyle 
f_{n,a,b}(z) \;\;=\;\; 
{{a+ z -1
\choose z}
{b+n-
z-1\choose  n- z}
\over  {a+b+n-1\choose n} }
\;\;=\;\;
{{-a\choose z} 
{-b\choose n-z }
\over  {-a-b\choose n} }
\quad \mbox{for $z=0,...,n$.}
\end {array}
\ea
The bivariate discrete distribution 
defined by the PDF
\ba \label {FOR_BIVARIATE_BETABINOMIAL_10}
\begin {array} {l}
\displaystyle 
f_{n_1, n_2,a,b} (z_1, z_2) \;\;=\;\;
{ {n_1\choose z_1} {n_2\choose z_2}
B(a+z_1+z_2, b+n_1+n_2-(z_1+z_2) ) \over B(a,b) }
\end {array}
\ea
for  $z_i=0,...,n_i$
is called {\em bivariate beta-binomial
distribution} 
with parameters $a,b \in (0;+\infty)$ and 
$n_1, n_2 \in \NAT$,
in short addressed as {\em bivariate 
beta-binomial distribution 
$BIBETA$-$Bi(n_1,n_2, a,b)$}.
Using the equations
(\ref {FOR_BINOMKOEFF_15}) and
(\ref {FOR_BETAFUNKTION_45}) provided
by the appendix 
\ref {Facts on binomial coefficients
and the beta function}, the PDF 
(\ref {FOR_BIVARIATE_BETABINOMIAL_10}) can be 
expressed as
\ba \label {FOR_BIVARIATE_BETABINOMIAL_20}
\begin {array} {l}
\displaystyle 
f_{n_1, n_2,a,b} (z_1, z_2)  \;\;=\;\; 
{ {n_1\choose z_1} {n_2\choose z_2}
\over {n_1+n_2\choose z_1+z_2} }
{{a+z_1+z_2-1 \choose z_1+z_2} {b+n_1-z_1+n_2-z_2 -1
\choose n_1+n_2-z_1-z_2}
\over {a+b+n_1+n_2-1 \choose n_1+n_2} }
\;\;=\;\; \\[4mm]
\displaystyle 
{ {n_1\choose z_1} {n_2\choose z_2}
\over {n_1+n_2\choose z_1+z_2} }
{{-a \choose z_1+z_2} {-b\choose n_1+n_2-z_1-z_2}
\over {-a-b\choose n_1+n_2} }
\quad \mbox{for $z_i=0,...,n_i$.}
\end {array}
\ea

\vspace {5mm}
\begin{proposition}
{\text{(\normalfont distributions under beta prior)} }
\label {PROPO_BIVARIATE_BETABINOMIAL_10}
Let $X$ have the beta distribution
$BETA(a,b)$ defined by the PDF
(\ref {FOR_BETA_10}).
Conditional under $X=x$, let
$Z_1$ and $Z_2$ be independent,
and let $Z_i$ have the
binomial distribution $Bi(n_i,y)$ with
$n_1, n_2\in \NAT$.
Then we have:
\begin {description}
\item [a)] The unconditional joint 
distribution of $Z_1$ and $Z_2$ is the 
bivariate 
beta-binomial distribution 
$BIBETA$-$Bi(n_1,n_2, a,b)$ with PDF 
provided by equations
(\ref {FOR_BIVARIATE_BETABINOMIAL_10})
and
(\ref {FOR_BIVARIATE_BETABINOMIAL_20}).
\item [b)] For $i=1,2$, the unconditional 
distribution of $Z_i$ is the univariate 
beta-binomial distribution 
$BETA$-$Bi(n_i, a,b)$, and the
conditional distribution of $X$
under $Z_i=z_i$ is the 
beta distribution
$BETA(a+z_i, b+n_i-z_i)$.

\item [c)] For $z_1=0,...,n_1$,
the conditional distribution of $Z_2$
under $Z_1=z_1$ is the 
univariate 
beta-binomial distribution 
$BETA$-$Bi(n_2, a+z_1,b+n_1-z_1)$
with PDF provided by equations
(\ref {FOR_BETABINOMIAL_10}) 
and (\ref {FOR_BETABINOMIAL_20}). 

\end {description} 
\end{proposition}

\paragraph{Proof of
of proposition
\ref{PROPO_BIVARIATE_BETABINOMIAL_10}.}
Consider assertion a).
Using the definition (\ref {FOR_BETA_20}) 
of the beta function $B(\cdot, \cdot )$ we obtain
$$ f_{Z_1, Z_2} (z_1, z_2) \;\;=\;\;
\int _{[0;1]} f_{Z_1, Z_2\vert Y=y} (z_1, z_2) 
f_Y(y) \, \mbox{d} y \;\;=\;\; $$
$$
\int _{[0;1]} {n_1\choose z_1}
{n_2\choose z_2} y^{z_1+z_2}
( 1-y )  ^{n_1+n_2- (z_1+z_2)}
{1\over B(a,b)} y^{a-1} (1-y)^{b-1} 
\, \mbox{d} y \;\;=\;\;
$$
$$ {1\over B(a,b)}  \int _{[0;1]} {n_1\choose z_1}
{n_2\choose z_2} y^{a+z_1+z_2-1}
( 1-y )  ^{b+n_1+n_2- (z_1+z_2)-1}
\, \mbox{d} y \;\;=\;\; $$
$$
{ {n_1\choose z_1} {n_2\choose z_2}
B(a+z_1+z_2, b+n_1+n_2-(z_1+z_2) ) \over B(a,b) }.$$
Comparing with equation
(\ref {FOR_BIVARIATE_BETABINOMIAL_10}) completes
the proof of assertion a).
Assertion b) is demonstrated analogously.
The assertion c) follows from assertions a) and b)
with standard formulae of probability
calculus.

\section
{Facts on binomial coefficients and the
beta function}
\label
{Facts on binomial coefficients
and the beta function}
For  $x\in \REL$ and integer $k$  
the binomial coeffizient $  
{x \choose k}$  is defined by
\ba \label{BINKOEFF}
{x \choose k} \quad := \quad \DREI
{0,\; \;}{\; \mbox{ falls } \; k <0,}
{1,\; \;}{\; \mbox{ falls } \; k=0,}
{\prod\limits_{i=1}^k { x - i+1 \over i} \, = \, 
{\prod _{i=1}^k ( x - i+1 )\over k!},
  \; \;}
{\; \mbox{ falls } \; k>0.} \ea
Binomial coefficients $y \choose k$ with 
$y < 0$ are called {\em negative binomial
coefficients}.
We have 
\ba \label{FOR_BINOMKOEFF_10}
{-x \choose k}  \;\; = \;\; (-1)^k {x+k-1 \choose k } \qquad \mbox{ 
for } x \in \REL \; \mbox{and integer $k$},
 \ea
and 
\ba \label{FOR_BINOMKOEFF_15}
{ {-x \choose k} {-y \choose n-k} \over
{-x -y \choose n } }
 \;\; = \;\; 
{ {x+k-1 \choose k } {y+(n-k)-1 \choose n-k }
\over {x+y +n-1 \choose n } }
\quad \mbox {for } x,y \in \REL,\, n,k \in \NATO. \ea
Using the well-known recursion
\ba \label {FOR_GAMMAFUNKTION_20}
\Gamma (x+1) \;\; = \;\; x\Gamma(x)
\quad \mbox{for }x\in (0;+\infty),
\ea
for the gamma function we can establish the
subsequent relations between the
beta function defined by 
(\ref {FOR_BETA_20}) and binomial coefficients:
\ba \label {FOR_BETAFUNKTION_45}
\begin {array}  {l}
\displaystyle 
B(s+k,t+m) \;\; =\;\; B(s,t)
{ {s+k-1\choose k} {t+m-1\choose m} 
\over {s+t+k+m-1\choose k+m} } {1\over
{m+k \choose m} } \;\; = \;\; 
\\[2mm]
\displaystyle 
B(s,t)
{ {-s\choose k} {-t\choose m} 
\over {-s-t\choose k+m} } {1\over
{m+k \choose m} } 
\end {array}
\ea
for $s,t >0$, $ k,m\in \NATO$.




\end{appendices}




\end{document}